# Heat Conduction and Thermal Switching Performance of Surface Plasmon Polaritons in Ag$_2$Se Quantum Dot Composite Polymer Film

*Congliang Huang[1, a)], Changkang Du[2], Qiangqiang Huang[1], Xiaodong Wang[3,b)]*

1. Research Center of Engineering Thermophysics, North China Electric Power University, Beijing 102206, China.
2. Department of Mechanical Engineering, The Hong Kong Polytechnic University, Hung Hom, Kowloon, Hong Kong SAR, China.
3. Technical Institute of Physics and Chemistry, Chinese Academy of Sciences, Beijing 100190, China.
a) Corresponding author E-mail: huangcl@ncepu.edu.cn.
b) Corresponding author E-mail: wangxd99@gmail.com.

**Abstract:** To stabilize the working temperature of an equipment, a solid-state thermal resistor is usually a requisite, which could adjust its heat conductance continuously according to the temperature. In this work, the thermal conductivity and the thermal switching performances of surface plasmon polaritons in the polymer films filled with Ag$_2$Se quantum dots (QDs) were theoretically analyzed, and a theoretical model was also derived to reveal the dependence of the thermal conductivity on the temperature and the structure of a composite film, which is verified to be effective by numerical calculations. It shows that the thermal conductivity will decrease following $\sim t^{-3}\exp(\zeta/Td)$ rule under the thin film limit, here $t$, $d$ and $T$ are film thickness, diameter of QDs and temperature, respectively, and $\zeta$ is a constant. A high thermal conductivity could be only realized at a device with a size lager than millimeter scale, due to the need of avoiding boundary scatterings of surface plasmon polaritons (SPPs). At the millimeter scale, the thermal conductivity could be reduced by 100 times by increasing temperature from 300 to 400 K, which suggests a very high thermal switching ratio almost in all kinds of solid-state thermal resistor. This study brings new insights in designing thermal resistor and understanding heat conduction in films by adjusting its structures.

# 1 Introduction

In the field of manipulating heat transport, many methods have been developed to tune the thermal conductivity by selecting proper nanostructures or materials [1,2], however these methods lack repeatable operability after the structure being selected [3]. The active thermal control technology (e.g., by electric or magnetic fields) has problems such as high energy consumption or/and limited life [4,5]. A solid-state thermal resistor is remaining a requisite, which could adjust its heat conductance continuously according to the (ambient) temperature, and thereby can stabilize the working temperature of an equipment. [6] Besides phonons and electrons as heat carriers in thermal resistor, surface polariton provides another pathway to modulate heat conduction [7]. However, the thermal conductivity of surface polaritons is usually not sensitive to temperature [7].

In recent years, optical properties of quantum dots (QDs) have attracted a wide interest and have been reported to be importantly dependent on the temperature besides their size [8-11]. This tunable optical property mainly originates from the changed permittivity, while the efficient temperature and size modulation of permittivity arises from tuned electron Fermi-Dirac statistics with the Fermi level playing a crucial role. [12] In this work, the possibility of tuning permittivity of QDs thus to modulate the thermal conductivity carried by surface plasmon polaritons (SPPs) in QD composite nanostructures will be focused, while there is still no such kind of works.

In the application of QDs, they are usually integrated into a matrix or host (such as polymers, chalcogenide glasses, oxides, or perovskites) [13,14]. By dispersing Ag2Se QDs in polymer films or organic solutions, experimental results confirm that its optical property is still temperature dependent. [15,16] In this work, we will theoretically discuss the heat conduction and thermal switching performance of polymer films filled with $Ag_2Se$ QDs, where the ultra-high molecular weight polyethylene (PE) polymer was applied as the matrix film.

# 2 Methodology

## 2.1 Thermal conductivity λ

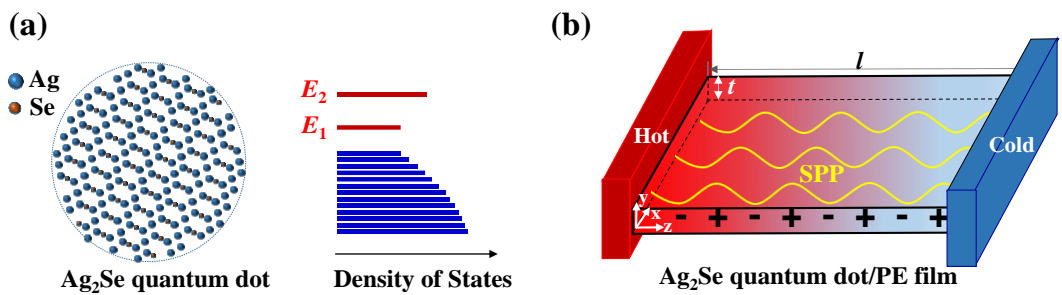

*Figure 1 Schematic of Ag2Se QD (a) and its PE composite film (b).*

The schematic of Ag$_2$Se QD and its electron density of states were shown in Figure 1a. For polymer films filled with Ag$_2$Se QDs and surrounded by air, as illustrated in Figure 1b, after heated up with a thermal bath, the electrons in QDs will oscillate and emit an electromagnetic field, which is called self-thermal emission. [17,18] This thermally excited field, while localized on the surface, are usually called SPPs. [19,20] Considering that SPPs could be only excited by electromagnetic waves in TM polarizations in a nonmagnetic media [7], only the TM evanescent component was considered here.

It has been proved by the fluctuation electrodynamics approach that the kinetic theory is applicative in a film thickness larger than 5 nm to predict the polariton thermal conductivity [21]. From the Boltzmann transport equation under the first-order relaxation time approximation, the thermal conductivity can be calculated by [22,23],

$$\lambda = \frac{1}{8\pi t} \int_0^\infty \hbar\omega \frac{\beta'}{\beta''} \left(\frac{\partial f_0}{\partial T}\right) d\omega, \qquad (1)$$

here $t$ is the film thickness, $\hbar$ is the reduced Plank constant, $\omega$ is the frequency, and $f_0$ is the equilibrium Bose-Einstein distribution, and $\beta$ is the wave vector along Z direction on the surface. Superscript " $'$ " and " $''$ " stand for the real and imaginary parts of a parameter in this work.

In Eq. (1), only $\beta'/\beta''$ depends on the material property (permittivity) of the film. $\beta'$ relates to the group velocity of SPPs, while $\beta''$ relates to the mean free path (MFP) of SPPs by $\Lambda = 1/2\beta''$ [24,25]. The dependence of $\beta$ on $\omega$ (dispersion relationship) could be determined by solving Maxwell equations under film boundary conditions. After solving the equations, the symmetric mode (even branch) of the relationship read as [26],

$$\tanh\left(\frac{tk_c}{2i}\right) + \frac{\varepsilon_c k_0}{\varepsilon_0 k_c} = 0, \qquad (2)$$

here, $k_j = \sqrt{(\varepsilon_j \omega/c)^2 - \beta^2}$ is the wave vector along the direction perpendicular to the film. Subscript $j$ = "0" and "c" indicates the air and the composite film respectively, $c$ is the speed of light in vacuum. Eq. (2) could be numerical resolved [22,23]. The dielectric function $\varepsilon_c$ of composite film was given in the following section.

## 2.2 Dielectric function

With the radii and the separation distances of QDs much smaller than the wavelength of the electromagnetic field, the QDs can be treated as point dipoles. [27,28] Under the dilute dispersion condition, the Maxwell-Garnett model based on the effective medium theory has been widely applied for evaluating dielectric function of a composite [29,30], $\varepsilon_c = \varepsilon_m\{\varepsilon_f + 2\varepsilon_m + 2\varphi[\varepsilon_f - \varepsilon_m]\}/\{\varepsilon_f + 2\varepsilon_m - \varphi[\varepsilon_f - \varepsilon_m]\}$, here $\varepsilon_f$ and $\varepsilon_m$ is the dielectric functions of the filler QDs and the matrix PE respectively. $\varepsilon_m = 2.3$ for PE matrix. [31,32] $\varphi$ is the filling ratio of QDs in the composite film.

For the QDs, the dielectric function can be described by a quasi-two-level electronic system, [33,34]

$$\varepsilon_f(\omega) = \varepsilon_\infty + \frac{f(T)\Omega\omega_{21}}{\omega^2 - \omega_{21}^2 - 2i\omega\gamma}. \tag{3}$$

Different from the classical Lorentzian oscillator, there is an additional factor $f(T)$ standing for the temperature-dependent electron population difference in QDs. $\varepsilon_\infty$ is the high frequency permittivity, and $\varepsilon_\infty = 4$ is a typical value for inorganic materials. $\gamma$ is the damping frequency. $\Omega$ is the oscillator strength, described by $\Omega = \Omega_0/d^3$, here $\Omega_0 = 12e^2|z_{12}|^2/\pi\hbar\varepsilon_0$. $z_{12}$ and $d$ is the atomic distance and diameter of QDs respectively. $\omega_{21}$ is the resonant frequency relating to the two energy levels. The relation between $\omega_{21}$ and the diameter ($d$) of QDs is built based on the k·p model as, [35]

$$\omega_{21} = \sqrt{\frac{E_g^2}{4\hbar^2} + E_p\frac{8.89^2}{3m_0 d^2}} - \sqrt{\frac{E_g^2}{4\hbar^2} + E_p\frac{4\pi^2}{3m_0 d^2}}, \tag{4}$$

here $E_g$ is the bandgap of the bulk material which equals 0.12 eV, $E_p$ is the Kane parameter which equals 8.87 eV, $m_0$ is the free electron mass. For a quasi-two-level electronic system, the expression of $f(T)$ was derived from quantum kinetic equation for electrons, [34]

$$f(T) = \frac{1}{\exp\left(\frac{\hbar\omega_{21}}{k_B T}\Delta\right)+1} - \frac{1}{\exp\left(\frac{\hbar\omega_{21}}{k_B T}(\Delta+1)\right)+1}, \tag{5}$$

where $\Delta = (E_1 - E_F)/\hbar\omega_{21}$. $k_B$ is the Boltzmann constant.

## 3 Theoretical analysis of λ

In this part, the dependence of λ on the temperature and structure of composite films was firstly analyzed for revealing the underlying co-effect dependency relationship. When the real part of the permittivity is much larger than the imaginary part (i. e., $\varepsilon_c' \gg \varepsilon_c''$) of the composite film, $\beta'/\beta''$ in Eq. (1) can be simplified as [36],

$$\frac{\beta'}{\beta''} = \frac{4}{(t\omega/c)^2} \cdot \frac{\varepsilon_c'^3}{\varepsilon_0^2(\varepsilon_c' - \varepsilon_0)} \cdot \frac{1}{\varepsilon_c''} + \frac{(\varepsilon_c' - \varepsilon_0)\varepsilon_c'}{2\varepsilon_0} \cdot \frac{1}{\varepsilon_c''}. \tag{6}$$

Given that the imaginary part of the permittivity of the PE film ($\varepsilon_m''$) is small enough to be neglected, the imaginary part of the permittivity of the composite film can be estimated by $\varepsilon_c'' = \varphi \cdot \varepsilon_f''$ in a dilute filling condition. [37] Therefore, Eq. (6) can be further written as,

$$\frac{\beta'}{\beta''} = \frac{4}{(t\omega/c)^2 \varphi} \cdot \frac{{\varepsilon_c'}^3}{{\varepsilon_0}^2(\varepsilon_c'-\varepsilon_0)} \cdot \frac{1}{\varepsilon_f''} + \frac{(\varepsilon_c'-\varepsilon_0)\varepsilon_c'}{2\varepsilon_0 \varphi} \cdot \frac{1}{\varepsilon_f''} , \qquad (7)$$

In this equation, $\varepsilon_f''$ can be derived from Eq. (3),

$$\frac{1}{\varepsilon_f''} = \frac{\omega_{21}^4 - 2\omega_{21}^2\omega^2 + 4\gamma^2\omega^2 + \omega^4}{2\gamma\Omega\omega_{21}\omega f(T)} , \qquad (8)$$

here, $\hbar\omega_{21}$ spans from 0.123-1.23 eV for QDs with diameters ranging from 1 to 10 nm. $\hbar\omega$ will be selected to be much smaller than 0.01 eV, while further including large values of $\omega$ in the integration of Eq. (1) will not further enlarge the integration obviously. Considering $\omega_{21} \gg \omega$, Eq. (8) can be simplified as,

$$\frac{1}{\varepsilon_f''} = \frac{\omega_{21}^3}{2\gamma\Omega\omega f(T)} . \qquad (9)$$

For the convenience of analysis, we define $X = \exp\left(\frac{\hbar\omega_{21}}{k_B T}\right)$, so Eq. (5) can be rewritten as,

$$\frac{1}{f(T)} = \frac{(X^\Delta+1)(X^{\Delta+1}+1)}{X^\Delta(X-1)} . \qquad (10)$$

It should be noted that $X \gg 1$ if $T < 500$ K, and $\Delta$ ranges from -1 to 1. According to the value of $X^\Delta$, Eq. (10) could be further simplified as,

$$\frac{1}{f(T)} \approx \begin{cases} X^\Delta, & \text{for } X^\Delta \gg 1 \quad (\Delta \to 1); \\ X^\Delta + 1, & \text{for } X^\Delta \sim 1 \quad (\Delta \to 0); \\ 1 + 1/X^{\Delta+1}, & \text{for } X^\Delta \ll 1 \quad (\Delta \to -1). \end{cases} \qquad (11)$$

For the requisite of a large regulation of thermal conductivity by changing temperature, $1/f(T)$ is needed to be sharply changed by changing X. It is obvious that the condition with $\Delta \to 1$ should be taken in Eq. (11), while $1/f(T) \approx 2$ for other two conditions. Under the condition $\Delta \to 1$, by substituting Eqs. (9) and (11) in Eq. (7), the following expression of $\beta'/\beta''$ could be taken,

$$\frac{\beta'}{\beta''} = \frac{a}{t^2\omega^3\varphi}\exp\left(\frac{\zeta}{Td}\right) + \frac{b}{\omega\varphi}\exp\left(\frac{\zeta}{Td}\right) , \qquad (12)$$

here $a = \frac{2\omega_0^3 c^2}{\Omega_0 \gamma} \frac{{\varepsilon_c'}^3}{{\varepsilon_0}^2(\varepsilon_c'-\varepsilon_0)}$, $b = \frac{\omega_0^3}{4\Omega_0 \gamma}\frac{(\varepsilon_c'-\varepsilon_0)\varepsilon_c'}{\varepsilon_0}$, and $\zeta = \hbar\omega_0\Delta/k_B$, are physical properties and will not depend on the nanostructure of the composite film. In deriving Eq. (12), $\Omega = \Omega_0/d^3$ and $\omega_{21} = \omega_0/d$ were applied, here $\omega_{21}$ is derived from Eq. (4). For Ag$_2$Se QDs, $\omega_0 = 1.87 \times 10^6$ m·Hz and $\zeta = 1.423 \times 10^{-5}\Delta$.

After substituting Eq. (12) in Eq. (1), the thermal conductivity can be further written as,

$$\lambda = \frac{a_1}{t^3\varphi}\exp\left(\frac{\zeta}{Td}\right)\int_0^\infty \frac{1}{\omega^3}d\omega + \frac{b_1}{t\varphi}\exp\left(\frac{\zeta}{Td}\right)\int_0^\infty \frac{1}{\omega}d\omega, \tag{13}$$

where $a_1 = \frac{k_B}{8\pi}a$ and $b_1 = \frac{k_B}{8\pi}b$. The approximation $e^\Theta/(e^\Theta - 1)^2 \approx \Theta^{-2}$ was applied in the derivation of Eq. (13), given that $\Theta = \frac{\hbar\omega}{k_B T} \ll 1$.

Eq. (13) clearly reveals the dependences of λ on film thickness (*t*), filling ratio of QDs (*φ*), diameters of QDs (*d*) and also temperature (*T*). Under the thin film limit (TFL) while $t \ll c/\omega$, the second part in Eq. (13) will be small enough to be neglected compared to the first part, which is the usual condition mostly discussed in previous works for obtaining a high thermal conductivity [22,38,39]. It should be noted that under $\Delta \to 1$, Eq. (13) is applicable to all composite films for a straightforward thermal conductivity estimation. $a_1$, $b_1$ and $\zeta$ are only determined by the electromagnetism properties of the composite films in Eq. (13). For other cases of Δ, Eq. (13) could be easily obtained by applying different expressions of Eq. (11), which will not be discussed in this work. It should be noted that there is an implicit assumption in Eq. (1) that MFP of SPPs should be much smaller than the length of the film. Otherwise, the ballistic transport characters of polaritons should be considered [40].

## 4 Results and discussion

For the Ag$_2$Se QDs, Ω is reported to be about 0.3 eV/ℏ ($1.5\omega_{21}$) at *d* = 6.2 nm. [34] From the value of Ω, $\Omega_0$ is calculated to be $1.087 \times 10^{-10}$ m$^3$/s, which will be used in the numerical calculation. $\hbar\gamma$ is about 0.1 eV for Ag$_2$Se QDs. *Δ* is selected to be 0.9 for satisfying $\Delta \to 1$, and *φ* is selected to be 0.1. Unless otherwise stated, the frequency in the integrand of Eq. (1) is selected to span from 5 to 50 Trad/s, corresponding to the frequencies that the optical phonons of Ag$_2$Se could have [41,42].

### 4.1 Dependences of λ on structure

Eq. (13) obviously shows that the thermal conductivity will firstly decrease by following $t^{-3}$ and then decrease by following $t^{-1}$, as illustrated in Figure 2(a), which has also been confirmed by numerical methods in Ref. [43]. This different dependence of film thickness suggests that the TFL is t < 1000 μm in this work. The inset of Figure 2(a) shows that 50 Trad/s is large enough to give a convergent thermal conductivity, validating our selection of 50 Trad/s as the integral upper limit in Eq. (1). The influences of frequencies on the thermal conductivity were shown in Figure 2(b). It shows that there is a $\omega^{-3}$ dependence of thermal conductivity spectra under the condition that *t* < 1000 μm, while the thermal conductivity spectra largely deviates from $\sim\omega^{-3}$ for *t* =

1000 μm. This also confirms the condition of TFL being that $t < 1000$ μm. In TFL condition (t < 1000 μm), the first part of Eq. (13) could give a precise result while the second part can be neglected.

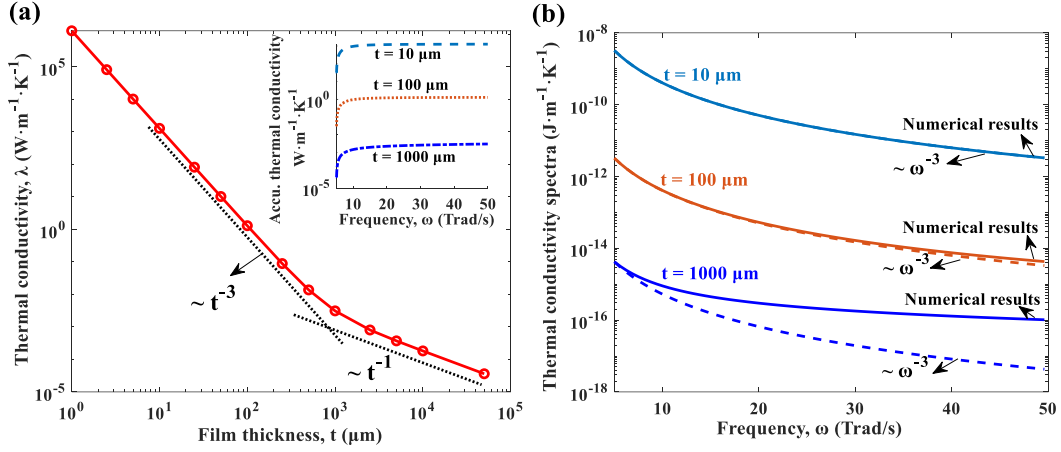

*Figure 2 Thermal conductivity of the film with d = 8nm and T= 300 K : (a) for different film thickness; (b) spectra at different ω. The solid and the dashed lines in (b) stand for the results obtained by numerical method and by the model with only first part (Eq. 13) respectively.*

The thermal conductivity at different temperatures was numerically calculated and shown in Figure 3(a). It shows that all numerical results for different films follow the same changing tendency of $\sim e^{\eta/T}$ as suggested by Eq. (13), here η = 1644.9, with a correlation coefficient larger than 0.999. The value of η calculated by our model is about 1600.9, which is very near to that obtained by numerical methods, indicating that our model could capture the changing tendency of λ well but with a small deviation of η. The thermal conductivity of films with different QD diameters were shown in Figure 3(b). With the decrease of QD diameters, the thermal conductivity follows the same rule $\sim e^{\xi/d}$ as suggested by Eq. (13), here ξ = 44.7, with a correlation coefficient larger than 0.999. The value of ξ calculated by Eq. (13) is about 42.7, which is also very near to that obtained by numerical methods.

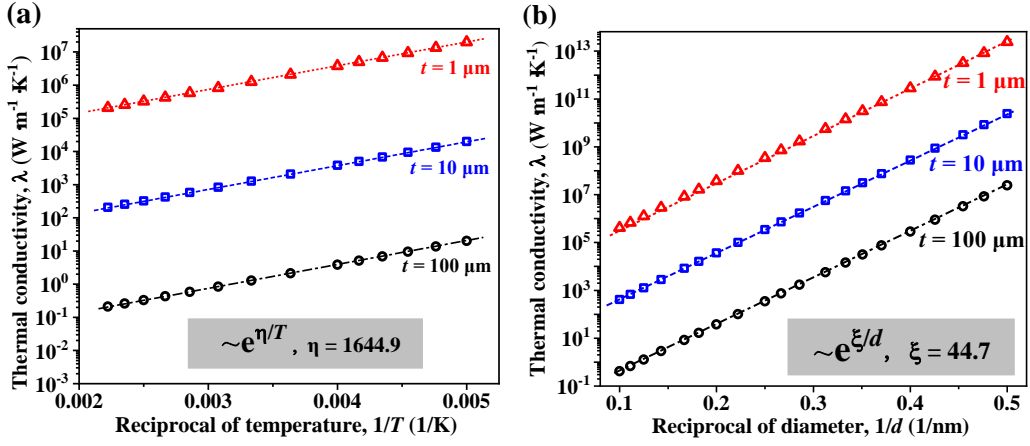

*Figure 3 Influences of temperatures and diameters of QDs on the thermal conductivity: (a) λ vs. 1/T for films filled with QDs of d = 8 nm; (b) λ vs. 1/d at T = 300 K. Triangles, squares, and circles represent numerical results of films with t = 1, 10, and 100 μm. Lines stand for fitting results.*

The real part of the vector β was numerically calculated and shown in Figure 4a, with the slope revealing the velocity of SPPs. It is clear that the velocity of SPPs for t < 100 μm is almost the speed of light in vacuum, while the velocity of SPP for t = 100 μm obviously deviates from the speed of light, confirming the condition of TFL being that $t < 1000$ μm. MFP of films with different thickness was shown in Figure 4b. It reveals that even for the 100-μm film which has a low λ of 1.28 W m$^{-1}$ K$^{-1}$, MFP should be still much larger than 1 meter. On the other hand, the size of microscale devices is much smaller than the meter scale, so MFP will be greatly reduced by the boundary scattering (size effect of device). This reduced λ of SPPs is small enough to be neglectable, thus obstructing the application of polaritons for enhancing heat dissipation in nanostructures [40,44]. This difficulty is also true for almost all polariton-contributed heat conduction, while the polariton is expected to be useful for heat dissipation at the nanoscale. [22,45]

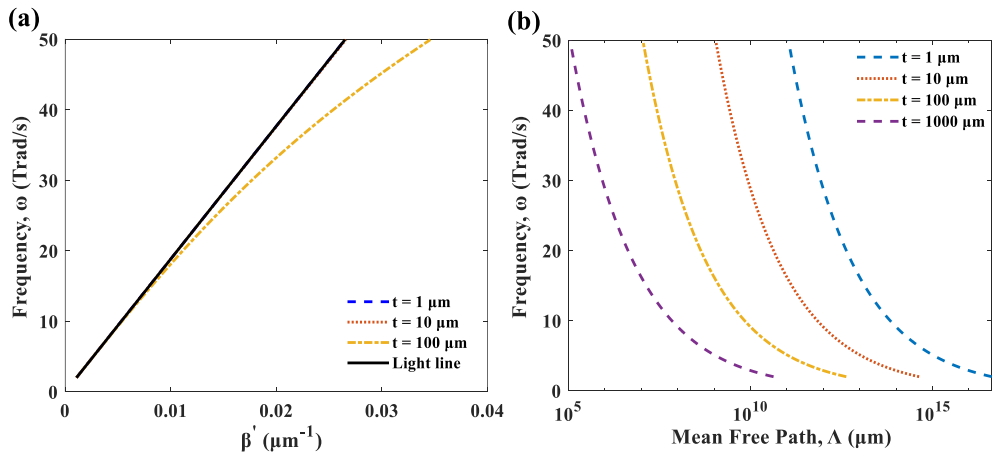

*Figure 4 Dispersion relation of SPPs in composite films filled with 8-nm-diameter Ag2Se QDs: (a) wave vector for different film thickness at T = 300 K; (b) mean free path for different film thickness at T = 300 K. The lines for t = 1 and 10 μm are overlapped by the light line in (a).*

### 4.2 Regulation of λ and thermal switching ratio

In the derivation of Eq. (12), we know that the value of $\beta'/\beta''$ reflects the MFP, while $\beta'$ is almost the speed of light. It can be seen that the second part in Eq. (12) will not depend on the film thickness *t*, while λ in Eq. (13) will increase with the decrease of *t*. If SPPs of high frequencies were excited, the first part in both Eqs. (12) and (13) will be small enough to be neglected and so the second part dominates. Under this condition, proper thickness could be selected to realize a relatively high λ while maintaining a relatively small MFP for application in small structures. In another words, this excitation of large frequencies will be helpful to overcome the drawback of the limited thermal conductivity caused by size effect in small structures. The excitation of high frequencies of SPPs in the film can be tuned by doping the film with special elements [46,47] or by coating hot and cold sources with special materials [48,49]. The excitation methods of high frequencies of SPPs will not be focused here, while more information can be found in Refs. [50,51]

Here, we will further probe the thermal conductivity carried by the SPPs with high frequencies. By applying a frequency span of 300-500 Trad/s, the thermal conductivity λ and MFP Λ were shown in Figure 5a and b respectively. The dashed line in Figure 5b which stands for Λ ≈ 1000 μm was further added in Figure 5a for comparison. Figure 5a shows that only the region circled out by the red line could satisfy a relatively high λ and simultaneously a relatively small Λ, which is needed for application in small structures. Here, λ will be about 0.1 W m$^{-1}$ K$^{-1}$ which is comparable to the thermal conductivity of polymer, while Λ is small enough in a small structure. It can be concluded that under special conditions, a relatively high thermal conductivity of SPPs could be realized for small structures but still not on nanostructures.

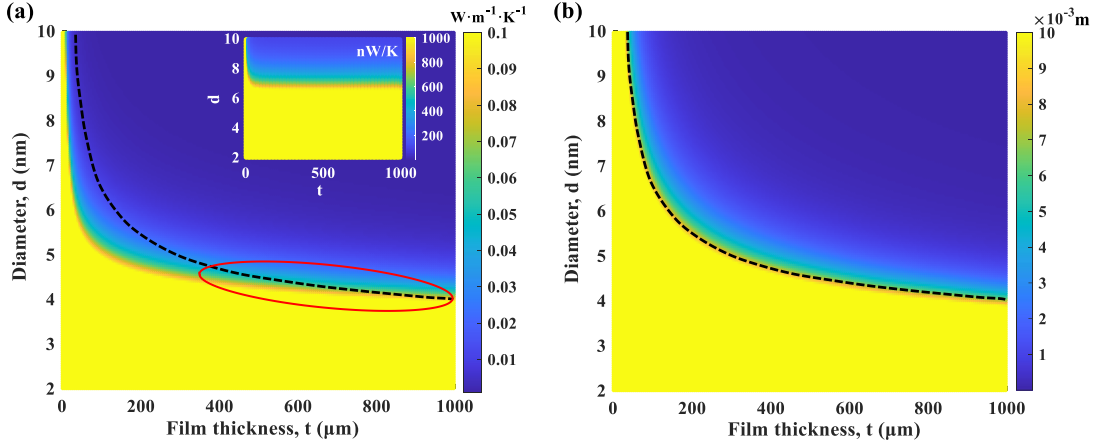

*Figure 5 Thermal conductivity with frequency spanning from 300 to 500 Trad/s (a) and mean free path of SPPs at 400 Trad/s (b). The inset figure in (a) shows the corresponding heat flux.*

The heat conduction of SPPs was shown in the inset figure of Figure 5a. It shows that this heat conduction will be on the 100 nW K$^{-1}$ level, which is on the same level as the phonon and also phonon-polariton heat conductions [52]. This suggests that the heat conduction of SPPs is large enough to be utilized in thermal switching of solid-state thermal resistor. In the region circled out in Figure 5a, taking the film with $d$ = 4.5 nm, $t$ = 400 μm as an example, numerical calculations show that: λ will decrease from 0.62 to 0.0058 W m$^{-1}$ K$^{-1}$ with temperature increasing from 300 to 400 K. The thermal conductivity could be decreased by 100 times by temperature, which suggests a very high thermal switching ratio almost in all kinds of solid-state thermal resistor [3].

## 5  Conclusions

In this work, to reveal the relationship between the thermal conductivity of SPPs and the structure of a composite film, a theoretical model was derived based on Boltzmann transport equation, which is verified by numerical method. The thermal conductivity will decrease following $\sim t^{-3}$ for a thin film, while decrease following $\sim t^{-1}$ when the thickness becomes larger than the thin film limit. The thermal conductivity also depends on the diameter of QDs and the temperature, following $\sim \exp(\zeta/Td)$, here ζ is a constant. It shows that a high thermal conductivity could be only realized at a device with a size lager than the millimeter scale, due to the need of avoiding boundary scatterings of SPPs. At the millimeter scale, the heat conduction of SPPs will be on the 100 nW K$^{-1}$ level, on the same magnitude scale as the phonon and also phonon-polariton heat conductions. At the millimeter scale, the thermal conductivity could be changed by 100 times by increasing temperature from 300 to 400 K, which suggests a high thermal switching ratio. This study brings new insights in designing thermal resistor and understanding heat conduction in films by adjusting its structures.

**Acknowledgement**

This work has been supported by the National Natural Science Foundation of China (No. 52322605).